\documentclass[12pt]{article}

\usepackage{epsfig}
 \begin{document} 

\def\be{\begin{equation}}
\def\ee{\end{equation}}
\def\ba{\begin{eqnarray}}
\def\ea{\end{eqnarray}}
\def\l{\label}
\def\n{\nonumber \\}
\def\b{\bibitem}

\title{Balance functions in  coalescence model}
  
\author{A.Bialas \\ M.Smoluchowski Institute of Physics \\Jagellonian
University, Cracow\thanks{Address: Reymonta 4, 30-059 Krakow, Poland;
e-mail:bialas@th.if.uj.edu.pl;}}
\maketitle

\begin{abstract}

It is shown that the quark-antiquark coalescence mechanism for pion
production allows to explain the small pseudorapidity
width of the balance function
observed for central collisions of heavy ions, provided effects of the
finite acceptance region and of the transverse flow are taken into
account. In contrast, the standard hadronic cluster model is not
compatible with this data.

\end{abstract}

{\bf 1.} Recently, measurements of balance functions \cite{isr} in
central collisions of heavy ions were reported by the STAR collaboration
\cite{star}. The striking feature observed in the data is the small
width of the balance functions (in rapidity and in pseudorapidity),
  as compared to the expectations from
the expanding thermally equilibrated quark-gluon plasma \cite{dp}. This
indicates that hadronization occurs only at the very late stage of the
development of the system \cite{pr}.

One may then ask if the hadronization properties of the system produced
in central collisions of heavy ions (as reflected in the balance
function) are similar to those observed in nucleon-nucleon collisions.
To investigate this problem we have evaluated the expected 
 width of the
balance function in pseudorapidity, using  the pion cluster
 model which some time ago was
successfully applied to nucleon-nucleon data \cite{f}\footnote{ Note
that in this model the quark-gluon structure of the system is entirely
ignored.}. Corrections due to the finite acceptance region of
\cite{star} and the effects of the transverse flow were included.

Our estimates show that to obtain consistency with the data of
\cite{star} for central collisions,
 the decay width of the pion cluster (in its rest frame) must
be substantially narrower than that corresponding to isotropic decay.
Thus one must conclude that the hadronization of the system produced in
heavy ion collisions is rather different from that produced in hadronic
collisions, where isotropic clusters can approximately account for the
data \cite{f}\footnote{This conclusion is not surprizing
since the measured rapidity width of the balance function in
nucleon-nucleon collisions \cite{isr} is about twice as large as that in
central heavy ion collisions \cite{star}. Our calculation shows that
neither finite acceptance nor transverse flow effects can account for
this difference.}.

Looking for a more adequate description, we considered the coalescence
model \cite{bud}, which we generalized to include correlations. We thus
assume that the hadronization proceeds in three steps. First, partons
form neutral clusters. Subsequently, each cluster decays into some
number of gluons and one\footnote{We consider just one pair for
simplicity. This is not essential for the conclusions.} $q\bar{q}$ pair
(either $u\bar{u}$ or $d\bar{d}$)\footnote{Throughout this paper by
quarks and antiquarks we always mean -in the spirit of the coalescence
model- the {\it constituent} quarks and antiquarks.}. Quarks and
antiquarks then recombine into positive, negative and neutral
pions. The remaining gluons form new clusters and the process is continuing
 until all partons are transformed into hadrons.

We show that this generalized coalescence model gives a good description
of the data from \cite{star}, provided the decay of clusters into quarks
and antiquarks is isotropic, which seems a rather natural assumption.
The obtained reduction of the width of the balance function (essential
to account for the data) is a natural consequence of the coalescence
process. It follows simply from the fact that the dispersion of the
average of two independent random variables is smaller than the
dispersion of each of them by factor $\sqrt{2}$.

It should be emphasized that we are discussing here only
 the angular distributions (expressed in terms of pseudorapidity).
 The natural assumption of approximately   isotropic 
 and uncorrelated cluster decay is then sufficient
 to describe the width of the balance function. This is not the case
  for rapidity distribution where more detailed information 
 on cluster decay is needed.

Our conclusion is that the generalized coalescence model provides a
natural explanation of the very narrow width of the balance function
observed in \cite{star}. This result is a consequence of a very general
fenomenon, characteristic for coalescence mechanism and thus, in our
opinion, it does not depend on details of the specific structure of
correlations between partons proposed in this note. Since, as we have
seen, the model based solely on hadronic degrees of freedom is not
adequate, we feel that our result provides a rather strong argument in
favour of the coalescence scenario.

{\bf 2.} The balance functions can be expressed in terms of the single
and double particle densities \cite{dp,jp}. Assuming, for simplicity,
the $(+-)$  symmetry we have
\ba
B(\Delta_2|\Delta_1)=D(-,\Delta_2|+,\Delta_1)-D(+,\Delta_2|+,\Delta_1)
\l{9} 
\ea 
with
\ba
D(+,\Delta_2|+,\Delta_1)=D(-,\Delta_2|-,\Delta_1)=
\frac{\int_{\Delta_2} d\eta_2 \int_{\Delta_1} d\eta_1
\frac{d^2n_{++}}{d\eta_{1}d\eta_{2}}} {\int_{\Delta_1} d\eta_{+} \frac{dn_+}{d\eta_+}}
\l{7}
\ea

\ba 
D(-,\Delta_2|+,\Delta_1)=D(+,\Delta_2|-,\Delta_1)=
\frac{\int_{\Delta_2} d\eta_2 \int_{\Delta_1} d\eta_1
\frac{d^2n_{+-}}{d\eta_{1}d\eta_{2}}} {\int_{\Delta_1} d\eta_{+}
\frac{dn_+}{d\eta_+}} \l{8} 
\ea
where  $dn/d\eta$ and $dn/d\eta_1d\eta_2$ are the corresponding 
particle densities in pseudorapidity.

The measurements of STAR require both particles to be in an
interval $-\Delta\leq \eta_1,\eta_2\leq \Delta$ (acceptance)
 while the difference of
(pseudo)rapidities is kept {\it fixed}. This suggests a change of
variables:
\ba
\eta_1-\eta_2 = \delta;\;\;\;\; (\eta_1+\eta_2)/2 =z. \l{ix}
\ea
The integrations must be performed over $dz$ with $\delta$ being fixed.
This implies that $z$ must
be kept inside the interval $-\bar{\Delta}\leq z\leq \bar{\Delta}$ where
$\bar{\Delta}= \Delta-|\delta|/2$. 

Thus the balance function measured in \cite{star} is given by
\ba
B_s(\delta;\Delta)= \frac{\int_{-\bar{\Delta}}^ {-\bar{\Delta}}dz \left[
d^2n_{+-}/dzd\delta -d^2n_{++}/dzd\delta\right]}
{\int_{-\Delta}^{+\Delta} d\eta_+ dn_+/d\eta_+}   \l{i5}
\ea

{\bf 3.} Consider first a model in which pions are produced in neutral,
isotropic clusters. It is well known that such models can account for
the gross features of the nucleon-nucleon data \cite{f}. The
distribution of clusters is denoted by $\rho_c(Y)$ where $Y$ is the 
cluster pseudorapidity.

To simplify the problem we assume  that clusters 
 decay into  two charged particles and any number of
neutrals. 

The distribution in cluster decay is 
\ba
\frac{dN_{+-}}{d\eta_+d\eta_-}= h(z-Y)f(\eta_+-Y)f(\eta_--Y)
\l{2}
\ea
where  
 $h(z-Y)$ is responsible for correlations: $h\equiv 1$ if the 
 decay products are uncorrelated.

The single particle distribution in cluster decay $\rho(\eta_+-Y)$
 is obtained by integration of (\ref{2}) over  rapidity
of one particle.
Both distributions  are normalized to unity.

The distribution of all particles is the convolution
\ba
\frac{dn_+}{d\eta_+}= \int dY \rho_c(Y) \rho(\eta_+-Y)  \l{3}
\ea
with identical formula for negative particles.

To evaluate the two-particle distributions one has to  take into account that
some pairs may come from different clusters and some others 
from one cluster. As is well-known (and can be easily confirmed 
by explicit calculation) the contribution from different
clusters cancels in the balance function and thus only $(+-)$ pairs from
one cluster do contribute. Their distribution is given by 

\ba
\frac{d^2n_{+-}}{d\eta_{+}d\eta_{-}}= 
\int dY \rho_c(Y) h(z-Y)f(\eta_+-Y)f(\eta_--Y) . \l{5}
\ea
Introducing  (\ref{5}) into (\ref{i5}) we have
\ba
B(\delta;\Delta)=\frac{ \int_{-\bar{\Delta}}^{+\bar{\Delta}} dz
\int dY \rho_c(Y) h(z-Y)f(\eta_+-Y)f(\eta_--Y)}
{\int_{-\Delta}^{\Delta} d\eta \int dY \rho(\eta-Y)}. \l{10a}
\ea

{\bf 4.} To continue, we  perform a simple exercise,
 assuming that all functions are
 Gaussians. We take\footnote{The normalization of 
the expression for $h(u)$ guarantees the
correct normalization in  (\ref{2}).}
\ba
\rho_c(Y)= \frac{<N_c>}{A\sqrt{\pi}} \exp(-Y^2/A^2);\;\;\; \n
f(u) =\frac1{f\sqrt{\pi}} \exp(-u^2/f^2);\;\;\;\;
h(u)= \sqrt{1+f^2/2g^2}\exp(-u^2/g^2)  \l{11}
\ea
Using  (\ref{11}) one can evaluate the single particle distribution in
cluster decay:
\ba
\rho(\eta)=\frac1{a\sqrt{\pi}}\exp(-\eta^2/a^2);\;\;\; a= f
\frac{\sqrt{1+f^2/4g^2}} {\sqrt{1+f^2/2g^2}}.  \l{i6}
\ea
This gives
\ba
\frac{dn_+}{d\eta_+}=\frac{<N_c>}{\sqrt{\pi(A^2+a^2)}}\exp[-\eta_+^2/(A^2+a^2)];\n
\int dY \rho_c(Y) h(z-Y)f(\eta_1-Y)f(\eta_2-Y)=\n= <N_c>
\frac1{f\sqrt{2\pi}} \exp(-\delta^2/2f^2) \frac1{F\sqrt{\pi}}
\exp(-z^2/F^2)   \l{i7}
\ea
with
\ba 
F^2 = \frac{f^2+2A^2[1+f^2/2g^2]}{2[1+f^2/2g^2]}   \l{i8}
\ea
This result  introduced into (\ref{10a}) allows to express the integrals
 over $z$ and $y$  in terms of the error function. The result is
\ba
B(\delta;\Delta)= \frac1{f\sqrt{2\pi}} \exp(-\delta^2/2f^2)
\frac{erf[(\Delta-|\delta|/2)/F]}{erf[\Delta/\sqrt{A^2+a^2}]}   \l{i9}
\ea
which completes the calculation.

This exercise shows that -in the limit of large acceptance- the width of
the balance function is determined by the parameter $f$ which describes
the cluster decay. 

{\bf 5.} These results can be used to evaluate  expectations
from isotropically decaying clusters which were found roughly
 compatible with
the data on hadron-hadron collisions \cite{f}. In this case the cluster 
decay distribution is
\ba
\rho(\eta) = \frac 1{2(\cosh \eta)^2}      \l{i10}
\ea
giving the cluster decay width $<|\eta|>=\log 2$. This can be approximated
by a Gaussian of the form (\ref{i6}) with $a = \log 2 \sqrt{\pi}\approx
1.23$. Ignoring for the time beeing the effect of finite acceptance, we
thus conclude from (\ref{i6}) that the expected width of the balance
function $<|\delta|>$ must be larger than $a\sqrt{2}/\sqrt{\pi} \approx
.98$ and thus by far exceeds the one measured in \cite{star}.

Finite acceptance [$\Delta =1.3$] of the STAR measurements \cite{star}
reduces the observed width of the balance function, as seen from
(\ref{i9}). 
This is not sufficient, however, to bring the data in
agreement with the model of isotropic pion clusters. The width
$<|\delta|>$ calculated from (\ref{i6}) for isotropic clusters with
$f/g=0$ (uncorrelated dcay) equals 0.67 at $A=3.5$ 
(this value of $A$ is roughly consistent
with data for central collisions \cite{br}), and does not change
significantly when $A$ varies around this value. Since, furthermore,
$<|\delta|>$ increases with increasing $f/g$, there is no chance to meet
the experimental value of $0.55$.

Another effect which may be responsible for the small width of the
balance function is the transverse flow. Indeed, the clusters which are
isotropic in their rest frame will not appear isotropic when moving with
a transverse velocity. As shown in \cite{z}, the distribution
(\ref{i10}) is then -to a good approximation- replaced by
\ba
\rho(\eta) = \frac1{2\cosh^2 \eta} \frac{\cosh Y_{\perp}}{[1+\sinh^2 Y_{\perp}
\tanh^2\eta]^{3/2}}  \l{i11}
\ea
where $Y_{\perp}$ is the transverse rapidity, $Y_{\perp}
= 0.5 \log [(1+v)/(1-v)]$,
 and $v$ is the tranverse velocity of the cluster.

%rys.1
\begin{figure}[htb]
\centerline{%
\epsfig{file=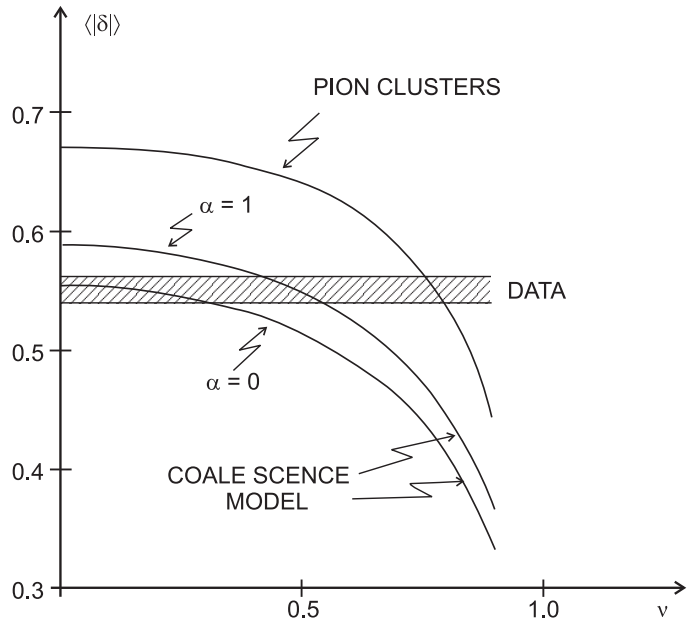,width=9cm}}
\caption{\small Width of  balance function versus velocity of transverse
flow.}
\end{figure}

We have calculated the width of the balance function with (\ref{i11})
approximated by a Gaussian\footnote{It was shown in \cite{z} that this
is a good approximation.} of the same width as that of (\ref{i11}). The
results are shown in Fig.1 where $<|\delta|>$ is plotted versus $v$, for
$A=3.5$. The measured values for most central events (as reported in
\cite{star}) are also indicated. One sees that the calculated width
decreases with increasing transverse velocity of clusters. 
One also sees that to obtain quantitative agreement with data the 
transverse velocity must approach $0.8 c$.
 Such a large value seems
difficult to reconcile with other estimates of the transverse velocity
\cite{ff}.

{\bf 6.} The pion cluster model discussed so far ignores entirely the
parton structure of the final system of hadrons. One may therefore not
be surprized that it fails to describe the data from central heavy ion
collisions. This argument suggests to try a model in which the parton
structure is built in from the beginning. To this end we investigated
 coalescence model \cite{bud} which we generalized to include the
correlations inside the system.
\smallskip

To introduce correlations we assume that -just before hadronization- the
QGP forms the weakly correlated neutral clusters. The clusters decay
into quarks$^4$, antiquarks and gluons. One cluster provides one$^3$
$q\bar{q}$ pair (either $u\bar{u}$ or $d\bar{d}$) and any number of
gluons. In the final step  quarks and antiquarks coalesce into
observed hadrons. The remaining gluons form again neutral clusters
and the process continues.

\smallskip
Thus the model we consider is basically the well-known coalescence model
\cite{bud} supplemented by a prescription for  correlations.
Since the coalescence model was rather succesful in description of
single particle spectra in central collisions of heavy ions
\cite{bud,bp}, it seems worthwhile to investigate its extention to
correlation phenomena (see also \cite{bb}). Admittedly, the proposed
extension is very simple - perhaps even simplistic. It contains, however,
all ingredients necessary to formulate and study the width of balance
functions which is of interest in this paper. Therefore we do not find
necessary at the moment to formulate and discuss a more general
and/or detailed approach.

{\bf 7.} To evaluate the balance function we need the distribution of
pairs of charged pions, same charge as well as opposite charge. The
pairs of same charge can be constructed by coalescence of the decay
products of four clus-\break 
\noindent ters (two U-clusters and two
D-clusters)\footnote{To shorten the wording, we call by U-cluster the
one decaying into $u\bar{u}$ and by D-cluster the one decaying into
$d\bar{d}$. Their distributions and decay properties are identical.}.
The distribution of the pairs of opposite charge consists of two terms:
one identical to the distribution of same charge pairs and another one,
arizing from coalescence of the decay products of two clusters (one
U-cluster and one D-cluster). Thus
the contributions involving four clusters
exactly cancel and we only
have to consider the distribution of pions of opposite charge which
result from coalescence of decay products of one U-cluster and one
D-cluster.

This distribution can be expressed as
\ba
\rho(\eta_+,\eta_-)=
\int dY_U dY_D \rho_G(Y_C,\Delta_Y)\n
\int d\eta_u d\eta_{\bar{u}}
f_q(\eta_u-Y_U)f_q(\eta_{\bar{u}}-Y_U)h_q([\eta_u+\eta_{\bar{u}}]/2-Y_U)\n
\int d\eta_d d\eta_{\bar{d}}
f_q(\eta_d-Y_D)f_q(\eta_{\bar{d}}-Y_D)h_q([\eta_d+\eta_{\bar{d}}]/2-Y_D)
\n\delta[\eta_+-(\eta_u+\eta_{\bar{d}})/2]\delta[\eta_--(\eta_d+\eta_{\bar{u}})/2]
\Phi[\eta_u-\eta_{\bar{d}}] \Phi[\eta_d-\eta_{\bar{u}}] \l{5.1}
\ea
where $f_q$ and $h_q$ are responsible for the distribution of quarks
in decay of either $U$ or $D$ cluster, while
 $\Phi$ summarizes the properties
of the coalescence process. Finally, $\rho_G(Y_C,\Delta_Y)$ denotes the joint
distribution of $U$ and $D$ clusters 
with average rapidity  $Y_C$, where
\ba
Y_C = \frac {Y_U+Y_D}2;\;\;\;\; \Delta_Y = Y_U-Y_D  \l{5.1a}
\ea
To simplify the discussion, in the following  we shall assume that 
$\rho_G$ factorizes:
\ba
\rho_G(Y_C,\Delta_Y)= \rho_C(Y_C) \rho(\Delta_Y)    \l{5.1b}
\ea

Taking advantage of the delta functions  
we can rewrite (\ref{5.1}) as
\ba
\rho(\eta_+,\eta_-)= 
\int dY_C d \Delta_Y \rho_G(Y_C,\Delta_Y) \int du_+ du_- \n
f_q\left(\eta_++\frac{u_+}2-Y_C\right)
f_q\left(\eta_-+\frac{u_-}2-Y_C\right)
h_q\left(z+\frac{u_++u_-}4-Y_C\right)\n
f_q\left(\eta_--\frac{u_-}2-Y_C\right)
f_q\left(\eta_+-\frac{u_+}2-Y_C\right)
h_q\left(z-\frac{u_++u_-}4-Y_C\right)\n
\Phi\left(u_++\Delta_Y\right)\Phi\left(u_-+\Delta_Y\right) 
 \l{5.2}
\ea
where $z= (\eta_++\eta_-)/2$.

To proceed, we again consider  Gaussians
\ba
f_q(x)=  \frac{1}{c\sqrt{\pi}} e^{-x^2/c^2};\;\;\;
h_q(x)= \sqrt{1+a^2/2h^2}\exp[-x^2/h^2]\n
\Phi(x) =\frac{1}{p\sqrt{\pi}} e^{-x^2/p^2}.    \l{5.4}
\ea

With this Ansatz, the integrals over $du_+du_-$ can be performed. The
result is 
\ba \rho(\eta_+,\eta_-)= C e^{-\delta^2/c^2} \int dY_C \rho_C(Y_C)
\exp\left[-\frac{4\left[z-Y_C\right]^2}{c^2}(1+c^2/2h^2)\right] \l{i14}
\ea 
where $\;\delta= \eta_+-\eta_-\;$ and $C$ is a constant, irreleveant for
further discussion.

The formula (\ref{i14}) can be now introduced into (\ref{i5}) and thus
the balance function can be calculated.
 In the limit of very large acceptance we  obtain

\ba
B_s(\delta;\Delta)_{\Delta\;\rightarrow \; \infty}= \frac1{c\sqrt{\pi}}
e^{-\delta^2/c^2}     \l{i17}
\ea
One sees that -in this limit- the width of the balance function depends
on one parameter which -to a large degree- determines also
 the distribution 
in decay of a cluster  into free quark and antiquark. Indeed, 
using (\ref{5.1}) and (\ref{5.4}), one can show that the
 decay distribution in the rest frame of the cluster is given by
\ba
\rho_p(\eta_u)=\frac1{d\sqrt{\pi}}
\exp\left[-\frac{\eta_u^2}{d^2}\right]
\l{i19}
\ea
where
\ba
d^2 = {c^2} \frac{1+c^2/4h^2}{1+c^2/2h^2}  .  \l{i19a}
\ea

{\bf 8.} It seems  natural to assume that  -in their rest frame- 
clusters decay isotropically.
 This means that their decay distribution is 
given by (\ref{i10}), the same as for the clusters of pions considered 
before. It follows that -for an uncorrelated decay ($c/h \approx 0$)- 
 the parameters $f$ in (\ref{11}) and $c$ in (\ref{5.4})
are identical.
 Comparing (\ref{i9}) and (\ref{i17}) we thus conclude that in the coalescence
model the width of the balance function is expected to be by factor
$\sqrt{2}$ smaller than that obtained for  pion clusters.
The reason is clear: the dispersion of the pion rapidity is
reduced by precisely this factor when the pion is formed by random
coalescence of a quark and an antiquark. 

Repeating the argument of the section 5, we thus conclude that -ignoring for
the moment the corrections for finite acceptance and effects of transverse
flow- the width of the balance function is expected to lie between .69
and .98 (the lower limit is obtained for $\alpha=(c/h)^2 = 0$, i.e. when
the decay products of a cluster are uncorrelated).

To compare this result with the data we  have to estimate the
corrections. To this end  we take the
Gaussian Ansatz for  $\rho_C$:
\ba 
\rho_C(Y_C)=\frac1{A\sqrt{\pi}}e^{-Y_C^2/A^2}   \l{i20}
\ea
which allows to evaluate explicitely the integrals in (\ref{i14}).
 We obtain
\ba
B_s(\delta;\Delta)= \frac1{c\sqrt{\pi}}
e^{-\delta^2/c^2} 
\frac{erf[2(\Delta-|\delta|/2)/b]}{erf[2\Delta/\sqrt{b^2+c^2}]}    \l{i21}
\ea
where
\ba
b^2= \frac{c^2+4A^2 (1+c^2/2h^2)}{1+c^2/2h^2}   \l{i22}
\ea

Using (\ref{i21}) one can now follow the argument of section 5 and
evaluate the width of the balance function, taking into account the
finite acceptance and the transverse flow. In Figure 1 the width of the
balance function evaluated from (\ref{i21}) is plotted versus $v$ for
$A=3.5$ and two values of the parameter $\alpha= c^2/h^2$. One sees that
these effects reduce substantially the calculated width. The value found
in \cite{star} for central collisions is reproduced with transverse
velocity below 0.5, consistent with other estimates of the
transverse flow \cite{ff}.

One also sees from the Fig. 1 that 
in the coalescence model the calculated width is smaller than
the value 0.65 found in \cite{star} for peripheral collisions. This is
not surprizing: in peripheral collisions a substantial part of the
particle production should resemble the elementary nucleon-nucleon
collisions which are not expected to follow the coalescence mechanism
\cite{bp} and are characterized by a significantly larger width of the
balance function \cite{isr,f}. As seen from Fig.1,
the width of the balance function 
calculated from the  pion cluster model 
(adequate for nucleon-nucleon collisions) is indeed close to 0.65.

{\bf 9.} In conclusion, we have shown that the coalescence mechanism
 implies a substantial reduction of the pseudorapidity width of the balance
function. This allows to explain the small width observed for central
collisions of heavy ions \cite{star}, provided the corrections due to
the finite acceptance region and to the transverse flow are taken into
account. This result  supports the coalescence
mechanism as the final stage of the process of hadronization.

\vspace{0.3cm}
{\bf Acknowledgements}
\vspace{0.3cm}

Thanks are due to W.Broniowski, W. Florkowski and 
K. Zalewski for comments and encouragement.
This investigation was supported in part by the by Subsydium of
Foundation for Polish Science NP 1/99 and by the Polish State Commitee
for Scientific Research (KBN) Grant No 2 P03 B 09322.

\end{document}